\documentclass[prd,aps,preprint,showpacs,nofootinbib,superscriptaddress]{revtex4}
\usepackage{epsfig}
\usepackage{amsmath}

\begin{document}


\title{Collins Fragmentation and the Single Transverse Spin Asymmetry}

\author{Feng Yuan}
\affiliation{Nuclear Science Division, Lawrence Berkeley National Laboratory, Berkeley, CA 94720}
\affiliation{RIKEN BNL Research Center, Building 510A, Brookhaven National Laboratory, Upton, NY 11973}
\author{Jian Zhou}
\affiliation{Nuclear Science Division, Lawrence Berkeley National Laboratory, Berkeley, CA 94720}
\affiliation{School of Physics, Shandong University, Jinan, Shandong 250100, China}

\begin{abstract}
We study the Collins mechanism for the single transverse spin
asymmetry in the collinear factorization approach. The correspondent
twist-three fragmentation function is identified. We show that the
Collins function calculated in this approach is universal.
We further examine its contribution to the single transverse spin
asymmetry of semi-inclusive hadron production in deep inelastic scattering and demonstrate that
the transverse momentum dependent and twist-three collinear approaches
are consistent in the intermediate transverse momentum region
where both apply.
\end{abstract}
\maketitle

\newcommand{\be}{\begin{equation}}
\newcommand{\ee}{\end{equation}}
\newcommand{\ben}{\[}
\newcommand{\een}{\]}
\newcommand{\beqn}{\begin{eqnarray}}
\newcommand{\eeqn}{\end{eqnarray}}
\newcommand{\Tr}{{\rm Tr} }

{\bf 1. Introduction.}
Single-transverse spin asymmetries (SSA) in hadronic processes have attracted
much attention from both experiment and theory sides, and great progress has been
made in the last few years.
They are defined as the spin asymmetries when we flip the transverse spin
of one of the hadrons in the scattering processes:
$A=(d\sigma(S_\perp)-d\sigma(-S_\perp))/(d\sigma(S_\perp)+d\sigma(-S_\perp))$
where $d\sigma$ is the differential cross section.
These physics involve nontrivial nucleon structure and strong interaction QCD dynamics.
In recent years, there has been great theoretical progress in exploring the underlying physics for
the SSAs observed in various hadronic processes~\cite{BroHwaSch02,Col02,{BelJiYua02},BoeMulPij03,JiMaYu04,{ColMet04}}. 
Especially, it has been found that the final/initial state interactions play very important 
roles to leading to a nonzero SSA in the Bjorken limit~\cite{BroHwaSch02}. These effects are
closely related to the gauge properties in the definition of the associated transverse
momentum dependent (TMD) parton distributions~\cite{Col02,{BelJiYua02},BoeMulPij03} and
the QCD factorizations for the relevant hadronic processes~\cite{JiMaYu04,{ColMet04}}. Based
on these developments, it has been shown~\cite{Ji:2006ub} that the two widely used approaches
to study SSA physics:
the transverse momentum dependent (TMD) approach~\cite{Siv90,{MulTan96}} and the twist-three
quark-gluon correlation in the collinear factorization approach~\cite{Efremov,qiu,{Eguchi:2006qz}}
are consistent in the intermediate transverse momentum
region where both apply. These progresses have laid solid
theoretical foundation to study QCD dynamics and the relevant nucleon
structure from the SSA phenomena.

However, this consistency has only been studied for the SSA contributions
coming from the polarized distributions of the incoming nucleon, where the
so-called Sivers function~\cite{Siv90} in the TMD approach is equivalent to the Qiu-Sterman
matrix element~\cite{qiu} in the twist-three collinear factorization approach~\cite{BoeMulPij03,{Ji:2006ub}}.
It has been difficult to extend to the SSAs associated with the fragmentation functions,
namely the Collins mechanism contribution to the SSA~\cite{Col93}.
The transverse momentum dependent Collins fragmentation function
describes the azimuthal hadron distribution correlated
with the quark transverse polarization vector~\cite{Col93}.
When combining with the quark
transversity distribution, it will generate the SSAs
in the semi-inclusive hadron production in deep inelastic scattering
(SIDIS)~\cite{Col93} and single inclusive hadron
production in $pp$ collisions~\cite{collins-s,{Ans94}}. It also contributes
to the azimuthal asymmetry in di-hadron production in
$e^+e^-$ annihilation process~\cite{Boer:1997mf}.
This contribution is very important not only because it is a significant contribution
to the SSA observables in hadronic processes, but also because its contribution is crucial
to extract the quark transversity distribution of nucleon, one of three leading twist
quark distributions~\cite{Jaffe:1991kp} which is weakly constrained~\cite{{Efremov:2006qm},Anselmino:2007fs}.
The experimental investigations of these physics haven been recently very active from both
SIDIS~\cite{hermes} and $e^+e^-$ processes~\cite{belle}.

Although they both belong to the naive-time-reversal-odd functions,
the Collins fragmentation function and the Sivers distribution have different
universality properties.
For example, the Sivers TMD quark distributions have opposite signs in the SIDIS
and Drell-Yan processes~\cite{Col02,{BroHwaSch02}}.
However, the TMD fragmentation functions are found to be universal
between different processes mentioned 
above~\cite{ColMet04,Metz:2002iz,collins-s,Gamberg:2008yt,{Meissner:2008yf}}. 
Especially, the final/initial state
interactions will not result into a sign change between different processes,
although they are important to retain the gauge invariance for the TMD fragmentation
functions~\cite{collins-s}.

Therefore, the previous studies on the consistency between the two approaches
for the distribution contributions to the SSAs are not straightforward to extend to the
fragmentation part, because the underlying physics and the roles played
by the initial/final state interactions are totally different~\cite{collins-s,{Metz:2002iz}}.
In particular, the twist-three quark-gluon correlation function in the twist-three
approach associated with the Collins contribution to the SSAs has not yet been identified.
In this paper we will study this issue.
There has been earlier attempt to construct the twist-three fragmentation
function~\cite{{Eguchi:2006qz},Koike:2002ti} contributing to the SSA in hadronic processes.
However, the function proposed there  has been shown to vanish
because of the universality arguments~\cite{Meissner:2008yf,{Gamberg:2008yt}}
(see also the discussions below).
In this paper, we will identify the twist-three fragmentation function corresponding to the
Collins function. We will further calculate the large transverse momentum
behavior of the Collins function from this twist-three fragmentation function, and
will find that it is universal. These results will be presented in Sec. 2.
In Sec.3, we will study the Collins contribution
to the SSA in SIDIS, and demonstrate that the TMD and collinear
factorization approaches are consistent in the intermediate transverse
momentum region. We conclude our paper in
Sec. 4.

{\bf 2. Collins Fragmentation at Large Transverse Momentum
and Twist-three Fragmentation Function.}
For the TMD quark fragmentation function, we define the
following matrix,
\begin{eqnarray}
  {\cal M}_h(z, p_{\perp}) &=&\frac{n^+}{z}
  \int \frac{d\xi^-}{2\pi}\frac{d^2 \xi_\perp}{(2\pi)^2}
  e^{-i(k^+\xi^--\vec{k}_\perp\cdot\vec{\xi}_\perp)}
 \label{ffdef} \\
  && \times \sum_X \frac{1}{3}\sum_a\langle 0|{\cal L}_{0}\psi_{\beta a}(0)
   |P_hX\rangle
 \nonumber \\ && \times \langle P_hX|(\overline{\psi}_{\alpha a}(\xi^-,\vec{\xi}_\perp) {\cal
L}_{\xi}^\dagger|0\rangle \nonumber \ ,
\end{eqnarray}
where $a=1,2,3$ is a color index, and $p_\perp$ is the transverse momentum of the
final state hadron with momentum $P_h$ relative to the fragmenting quark $k$.
The quark momentum $k$ is dominated by
its plus component $k^+=\frac{1}{\sqrt{2}}(k^0+k^z)$, and we have
$P_h^+=zk^+$ and $\vec{k}_\perp=-\vec{p}_\perp/z$. For convenience, we have chosen
a vector $n=(1^+,0^-,0_\perp)$ which is along the plus momentum direction.
The gauge link ${\cal L}_\xi$ is along the direction $v$ conjugate to $n$. In the case we need
to regulate the light-cone singularity, we will
choose an off-light-cone vector $v=(v^+,v^-,0_\perp)$ with $v^-\gg v^+$
and further define $\hat \zeta^2=(v\cdot P_h)^2/v^2$~\cite{JiMaYu04}.
The leading order expansion of the
above matrix leads to two fragmentation functions for a scalar meson,
\begin{eqnarray}
{\cal M}_h&=&\frac{1}{2}\left[D(z,p_\perp)\not\!
n+\frac{1}{M}H_1^\perp(z,p_\perp)\sigma^{\mu\nu}p_{\mu\perp}
n_\nu \right] \ , \label{tmdff}
\end{eqnarray}
where $M$ is a mass scale chosen for convenience, and the second term
defines the Collins function $H_1^\perp$.
From the above equation, we can further define the transverse-momentum
moment of the Collins function~\cite{MulTan96}:
$ \hat H(z)=\int{d^2p_\perp}\frac{p_\perp^2}{2M} H_1^\perp(z,p_\perp)$.
By integrating out the transverse momentum, the fragmentation function will only
depend on the longitudinal momentum fraction $z$ of the quark carried
by the final state hadron.
It is straightforward to show that this function can be written as a twist-three matrix
element of the fragmentation function,
\begin{eqnarray}
\hat H(z)&=&{n^+}{z^2}\int\frac{d\xi^-}{2\pi}e^{ik^+\xi^-}\frac{1}{2}\left\{
{\rm Tr}\sigma^{\alpha +}\langle 0|\left[iD_\perp^\alpha+\int_{\xi^-}^{+\infty} d\zeta^- gF^{\alpha+}(\zeta^-)\right]
\psi(\xi)|P_hX\rangle\right.\nonumber\\
&&\times\left.\langle P_hX|\bar\psi(0)|0\rangle+h.c.\right\} \ ,
\end{eqnarray}
where we have chosen the gauge link in Eq.~(1) goes to $+\infty$, and
$F^{\mu\nu}$ is the gluon field strength tensor and we have suppressed
the gauge links between different fields and other indices for simplicity.
Since the Collins function is the same under different gauge links~\cite{Metz:2002iz,collins-s,Gamberg:2008yt,{Meissner:2008yf}},
we shall obtain the same result if we replace $+\infty$
by $-\infty$ in the above equation. This will immediately show that
the matrix element used in~\cite{Koike:2002ti} vanishes because of the universality
property of the Collins fragmentation function~\cite{Meissner:2008yf}. From the above
definition, we can see that $\hat H(z)$ involves derivative on the
quark field and the filed strength tensor explicitly. Therefore, it belongs
to more general twist-three fragmentation functions~\cite{Ji:1993vw}.
For example, extending the above definition, we can define a two-variable
dependent twist-three fragmentation function as,
\begin{eqnarray}
\hat H_D(z_1,z_2)&=&{n^+}{z_1z_2}\int\frac{d\xi^-d\zeta^-}{(2\pi)^2}e^{ik_2^+\xi^-}e^{ik_g^+\zeta^-}
\frac{1}{2}\left\{{\rm Tr }\sigma^{\alpha +}\langle 0|iD_\perp^\alpha(\zeta^-)\psi(\xi^-)
|P_hX\rangle\right.\nonumber\\
&&\times\left.\langle P_hX|\bar\psi(0)|0\rangle+h.c.\right\} \ ,
\end{eqnarray}
where $k_i^+=P^+/z_i$ and $k_g^+=k_1^+-k_2^+$. Similarly,
we can define a $F$-type fragmentation function by replacing $D_\perp^\alpha$
with $F^{+\alpha}$. However, the $F$ and $D$ types are
related to each other by using the equation of motion~\cite{Ellis:1982wd}.
For our case, it is easy to show that~\cite{Eguchi:2006qz},
\begin{equation}
\hat H_D(z_1,z_2)=PV\left(\frac{1}{\frac{1}{z_1}-\frac{1}{z_2}}\right) \hat H_F(z_1,z_2)+
\delta\left(\frac{1}{z_1}-\frac{1}{z_2}\right) \hat H(z_1) \ ,
\end{equation}
where $PV$ stands for the principal value.
Therefore, they are not independent. In the following calculations we will
only keep $\hat H_F$ and $\hat H$ in the final results.

The above $\hat H_D$ function is different from the twist-three fragmentation
function $\hat E(z_1,z_2)$ introduced in~\cite{Ji:1993vw}.
In particular, $\hat H_D$ is the imaginary part whereas $\hat E$
is the real part of the same matrix element involving twist-three quark fragmentation
functions. Explicitly, if we replace $iD_\perp$ with $D_\perp$ in Eq.~(5) we will
obtain the definition of $\hat E$. If the time-reversal-invariance argument applies
to the fragmentation functions the above $\hat H_D$ function would vanish.
However, this argument does not apply here~\cite{Col93},
such that the $\hat H_D$ function exists
and contributes to the SSA in hadronic process. We emphasize that
it is actually this
function which corresponds to the Collins mechanism.

\begin{figure}[t]
\begin{center}
\includegraphics[width=9cm]{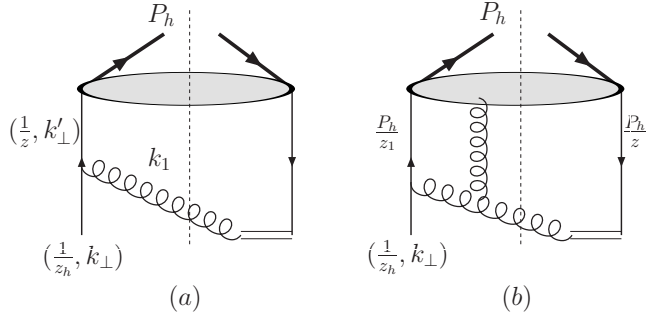}
\end{center}
\vskip -0.4cm \caption{\it Typical Feynman diagrams for the transverse momentum
dependent Collins fragmentation function calculated
in the collinear factorization approach: contributions
from (a) $\partial_\perp$ and (b) $A_\perp$ associated
operators in the  twist-three quark-gluon correlation functions.}
\end{figure}

Therefore, the above defined $\hat H_D$ ($\hat H_F$) and $\hat H$ Eqs.~(4,5)
will be our starting point to formulate the Collins mechanism in the collinear
factorization approach. First, we can calculate the transverse momentum
dependence of the Collins function in the perturbative region from the twist-three
fragmentation functions $\hat H_D$ and $\hat H$. To do this, we will have
to not only calculate the perturbative diagrams with gluon radiation,
but also to perform the twist expansion and take into
account full contributions from the $\partial_\perp$ and $A_\perp$
operators in the definitions of $\hat H_D$ and $\hat H$ at this
order~\cite{Ellis:1982wd}. We plot the typical Feynman diagrams in Fig.~1 for
the Collins function calculation from these contributions,
where a transversely polarized quark (with momentum
$k=P_h/z_h+k_\perp$) fragments into a final state hadron $P_h$ by radiating
a gluon with momentum $k_1$. For  the contribution from Fig.~1(a), we do
collinear expansion of the partonic scattering amplitude
in terms of $k_\perp'$, the transverse momentum
of the quark which couples to the final state hadron as shown in Fig.~1(a).
Combining this collinear expansion with the hadron fragmentation matrix
will form the $\partial_\perp\psi$ associated correlation function in Eqs.~(4,5)~\cite{Ji:1993vw}.
Similarly, the contributions from Fig.~1(b) with transverse gluon field $A_\perp$ connecting
the partonic part and the hadron fragmentation part will result
into the $A_\perp$ associated correlation function.
These contributions have to be sorted into
gauge invariant functions such as $\hat H_F$ ($\hat H_D$) and $\hat H$, respectively.
While the detailed derivations will be presented in a forthcoming publication,
here we summarize the final result,
\begin{eqnarray}
H_1^\perp(z_h,p_\perp)&=&\frac{\alpha_s}{2\pi^2}\frac{2M}{(p_\perp^2)^2}
\int \frac{dz}{z}\left[A+\delta(\hat\xi-1)\hat H(z)C_F\ln\frac{\hat\zeta^2}{p_\perp^2}\right] \ ,
\end{eqnarray}
where $\hat \xi=z_h/z$ and the function $A$ is defined as
\begin{eqnarray}
A&=&C_F\left[\left(z^3\frac{\partial}{\partial z}\frac{\hat H(z)}{z^2}\right)(-2\hat\xi)+
{\hat H(z)} \frac{2\hat\xi^2}{(1-\hat\xi)_+}\right]+\int \frac{dz_1}{z_1^2}
PV\left(\frac{1}{\frac{1}{z}-\frac{1}{z_1}}\right)\hat H_F(z,z_1)\nonumber\\
&&\times
\left[-C_F\frac{2z_h}{z}\left(1+\frac{z_h}{z_1}-\frac{z_h}{z}\right)-\frac{C_A}{2}\frac{2z_h}{z}
\frac{zz_1(z+z_1)-z_h(z^2+z_1^2)}{z(z-z_1)(z_1-z_h)}\right]\ .
\end{eqnarray}
In the above calculations, we have adopted an off-light-cone gauge link in Eq.~(1)
to regulate the light-cone singularity and $\hat\zeta^2$ has been defined above.

An important check of the above result is its universality property. Indeed,
we find that our calculations are free of the gauge link direction
used in Eq.~(1), i.e., the Collins function in Eq.~(6) is universal.
In particular, in the calculations we find that the gauge link does not
contribute to a pole in the Feynman diagrams of Fig.~1. Therefore,
the gauge links going to $+\infty$ and $-\infty$ lead to the same results. This is
consistent with the universality argument for the Collins fragmentation
function~\cite{ColMet04,{collins-s}}. We have checked that the contribution
from the twist-three function $\hat E_F$ introduced in~\cite{Koike:2002ti}
is also consistent with the universality property as those calculated above.

{\bf 3. Collins Effect in Semi-inclusive DIS.}
In this section, we extend to calculate the
the Collins contribution to the SSA in semi-inclusive DIS, $e p_\uparrow\rightarrow e' \pi X $,
and show that the TMD and collinear factorization approaches are consistent
in the intermediate transverse momentum region $\Lambda_{\rm QCD}\ll P_{h\perp}\ll Q$,
where $\Lambda_{\rm QCD}$ is the typical nonperturbative scale and $P_{h\perp}$ is the
transverse momentum of the final state hadron. Again, the
above defined $\hat H_D$ and $\hat H$ will be our starting
basis to calculate this contribution in the collinear factorization
approach.

In the SIDIS process $ e p_\uparrow\rightarrow e' \pi X $ where the incoming
nucleon is transversely polarized, the transverse spin
dependent differential cross section can be formulated as
\begin{eqnarray}
\frac{d\sigma (S_{\perp})}{dx_B dy dz_h d^2\vec{P}_{h\perp} }
=\frac{2 \pi \alpha_{em}^2}{Q^2} L^{\mu \nu }(\ell,q) W_{\mu \nu}(P_A,S_{\perp},q,P_h)\ ,
\end{eqnarray}
where $\alpha_{em}$ is the electromagnetic coupling,
$\ell$ and $P_A$ are incoming momenta for the lepton and nucleon, $S_\perp$
the polarization vector for nucleon,
$q$ the momentum for the exchanged virtual photon with $Q^2=-q.q$,
$P_h$ is the momentum for the final state hadron. The kinematic variables are defined
as $x_B=\frac{Q^2}{2 P_A \cdot q}$, $z_h=\frac{P_A \cdot P_h}{P_A \cdot q}$,
$y=\frac{P_A \cdot q}{P_A \cdot \ell}$. $L^{\mu\nu}$ and $W^{\mu\nu}$ are
leptonic and hadronic tensors, respectively.
To calculate the above differential cross section,
it is convenient to decompose the hadronic tensor into several terms:
$W^{\mu\nu}=\sum_i W_iV_i^{\mu\nu}$, where $V_i$ follow the definitions
of~\cite{meng}.

\begin{figure}[t]
\begin{center}
\includegraphics[width=12cm]{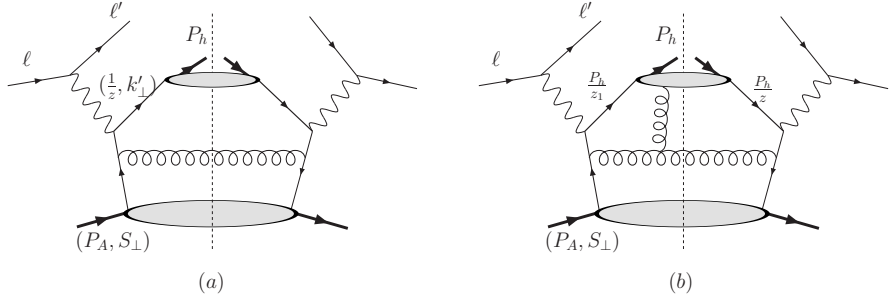}
\end{center}
\vskip -0.4cm \caption{\it Typical Feynman diagrams for the Collins
mechanism contributions to the single spin asymmetry in
semi-inclusive deep inelastic scattering. Again, we will
have contributions from (a) $\partial_\perp$ and (b) $A_\perp$ associated
operators in the  twist-three quark-gluon correlation functions.}
\end{figure}

As mentioned above, we will calculate the transverse spin dependent differential
cross section from the Collins mechanism, in particular the contributions from the twist-three
fragmentation functions $\hat H_D$ and $\hat H$ defined in Sec.2.
We follow the same procedure as that for the large transverse momentum Collins
fragmentation function calculated in the last section, and we will take into account the
contributions from both $\partial_\perp$ and $A_\perp$ associated fragmentation
matrix elements in Eqs.~(4,5). The relevant Feynman
diagrams can be drawn accordingly, and we show two examples
in Fig.~2. Similarly, their contributions can be
summarized into the terms associated with the gauge invariant functions
$\hat H_F$ ($\hat H_D$) and $\hat H$.
Furthermore, we are interested in the differential cross
section in the intermediate transverse momentum region
$\Lambda_{\rm QCD}\ll P_{h\perp}\ll Q$.
In the limit of $P_{h\perp}\ll Q$, we find that only $V_4$ and $V_9$ in the hadronic
tensor decomposition contribute in the leading power of $P_{h\perp}/Q$.
These two terms contribute the same to the differential cross sections
except the azimuthal angular dependence: the contribution from $V_4$
is proportional to $\cos(2\phi_h)\sin(\phi_s-\phi_h)$ whereas that from $V_9$ is
proportional to $\sin(2\phi_h)\cos(\phi_s-\phi_h)$, where $\phi_h $
and $\phi_s$ are the azimuthal angles of the transverse momentum $P_{h\perp}$ and
the polarization vector $S_{\perp}$ relative
to the lepton scattering plane.
The total contributions from these two terms will be proportional to $\sin(\phi_h+\phi_s)$,
\begin{eqnarray}
{\frac{d \sigma(S_{\perp})}{dx_B dy dz_h d^2 \vec{P}_{h\perp}}}\Big|_{P_{h\perp}\ll Q}^{V_4+V_9}&=&
\frac{4 \pi \alpha^2_{em} s}{Q^4} x_B(1-y) \sin(\phi_h+\phi_{s})
\frac{1}{z_h^2} \frac{\alpha_s}{2 \pi^2}
\frac{1}{ |\vec{q}_\perp|^3}
\nonumber \\
&&
\int \frac{dxdz}{xz}h_1(x)
\left \{ A \delta({\xi} -1)+ B \delta(\hat{\xi}-1) \right \},
\end{eqnarray}
in the limit of $P_{h\perp}\ll Q$, where $\xi=x_B/x$ and $\hat \xi=z_h/z$ and $A$ function
has been given in Eq.~(7) and $B$ is defined as
\begin{eqnarray}
B=C_F \hat H(z_h) \left [ \frac{2{\xi}}{(1-{\xi})_+}
+2 \delta ({\xi} -1) \ln \frac{Q^2}{\vec{q}_\perp^{\ 2}} \right ] .
\end{eqnarray}
Following the same procedure as that in~\cite{Ji:2006ub} for the Sivers effects, we will
find that the above single transverse spin dependent differential
cross section calculated from the twist-three fragmentation functions
$\hat H_D$ and $\hat H$ in the collinear factorization approach can be reproduced
by the TMD factorization for the same observable~\cite{JiMaYu04} by using the
large transverse momentum Collins fragmentation function calculated
in Sec.2, and the known results for the quark transversity distribution
and the soft factor~\cite{Ji:2006ub}.
This clearly demonstrates that in the intermediate transverse momentum
region, the twist-three collinear factorization approach and the
TMD factorization approach provide a unique picture for the Collins
contribution to the SSA in the semi-inclusive DIS.

{\bf 4. Conclusion.}
In this paper, we have studied the Collins mechanism contribution
to the single spin asymmetry in semi-inclusive hadron production in DIS
process. We have identified the corresponding twist-three fragmentation
function, and shown that the transverse momentum dependent
and collinear factorization approaches are consistent in the intermediate
transverse momentum region. Especially, we have also demonstrated that
the Collins fragmentation function calculated is universal, and free of
the gauge link direction. It will be important
to extend our calculations to the Collins contributions to the SSAs in other
processes, such as in hadron production in polarized
$pp$ scattering and di-hadron correlation in $e^+e^-$ annihilation. We will
address these issues in a future publication, together with a detailed derivation
of this paper.

We thank Bowen Xiao for the collaboration at the early stage of this work.
We are also grateful to Yuji Koike, Andreas Metz, and Werner Vogelsang for the 
discussions and comments.
This work was supported in part by the U.S. Department of Energy
under contract DE-AC02-05CH11231. We are grateful to RIKEN,
Brookhaven National Laboratory and the U.S. Department of Energy
(contract number DE-AC02-98CH10886) for providing the facilities
essential for the completion of this work.


\begin{thebibliography}
\frenchspacing

\bibitem{BroHwaSch02}
S.~J.~Brodsky, D.~S.~Hwang and I.~Schmidt,
Phys.\ Lett.\ B {\bf 530}, 99 (2002);
Nucl.\ Phys.\ B {\bf 642}, 344 (2002).

\bibitem{Col02}
J.~C.~Collins,
Phys.\ Lett.\ B {\bf 536}, 43 (2002).

\bibitem{BelJiYua02}
X.~Ji and F.~Yuan,
Phys.\ Lett.\ B {\bf 543}, 66 (2002);
A.~V.~Belitsky, X.~Ji and F.~Yuan,
Nucl.\ Phys.\ B {\bf 656}, 165 (2003).

\bibitem{BoeMulPij03}
D.~Boer, P.~J.~Mulders and F.~Pijlman,
Nucl.\ Phys.\ B {\bf 667}, 201 (2003).

\bibitem{JiMaYu04}
  X.~Ji, J.~P.~Ma and F.~Yuan,
  Phys.\ Rev.\ D {\bf 71}, 034005 (2005);
Phys.\ Lett.\ B {\bf 597}, 299 (2004).

\bibitem{ColMet04}
J.~C.~Collins and A.~Metz,
Phys.\ Rev.\ Lett.\  {\bf 93}, 252001 (2004).

\bibitem{Ji:2006ub}
  X.~Ji, J.~W.~Qiu, W.~Vogelsang and F.~Yuan,
  Phys.\ Rev.\ Lett.\  {\bf 97}, 082002 (2006);
  Phys.\ Rev.\  D {\bf 73}, 094017 (2006);
  Phys.\ Lett.\  B {\bf 638}, 178 (2006);
  Y.~Koike, W.~Vogelsang and F.~Yuan,
  Phys.\ Lett.\  B {\bf 659}, 878 (2008).


\bibitem{Siv90}
D.~W.~Sivers,
Phys.\ Rev.\ D {\bf 41}, 83 (1990);
Phys.\ Rev.\ D {\bf 43}, 261 (1991).

\bibitem{MulTan96}
P.~J.~Mulders and R.~D.~Tangerman,
Nucl.\ Phys.\ B {\bf 461}, 197 (1996) [Erratum-ibid.\ B {\bf 484},
538 (1997)];
D.~Boer and P.~J.~Mulders,
Phys.\ Rev.\ D {\bf 57}, 5780 (1998).


\bibitem{Efremov}
  A.~V.~Efremov and O.~V.~Teryaev,
  Sov.\ J.\ Nucl.\ Phys.\  {\bf 36}, 140 (1982);
  Phys.\ Lett.\ B {\bf 150}, 383 (1985).

\bibitem{qiu}
J.~Qiu and G.~Sterman,
Phys.\ Rev.\ Lett.\  {\bf 67}, 2264 (1991);
  Nucl.\ Phys.\ B {\bf 378}, 52 (1992);
Phys.\ Rev.\ D {\bf 59}, 014004 (1999);
  C.~Kouvaris, J.~W.~Qiu, W.~Vogelsang and F.~Yuan,
  Phys.\ Rev.\  D {\bf 74}, 114013 (2006).

\bibitem{Eguchi:2006qz}
  H.~Eguchi, Y.~Koike and K.~Tanaka,
  Nucl.\ Phys.\  B {\bf 752}, 1 (2006);
  Nucl.\ Phys.\  B {\bf 763}, 198 (2007).



\bibitem{Col93}
J.~C.~Collins,
Nucl.\ Phys.\ B {\bf 396}, 161 (1993).


\bibitem{Ans94}
M.~Anselmino, M.~Boglione and F.~Murgia,
Phys.\ Lett.\ B {\bf 362}, 164 (1995);
M.~Anselmino and F.~Murgia,
Phys.\ Lett.\ B {\bf 442}, 470 (1998).

\bibitem{collins-s}
  F.~Yuan,
  Phys.\ Rev.\ Lett.\  {\bf 100}, 032003 (2008);
  Phys.\ Rev.\  D {\bf 77}, 074019 (2008).

\bibitem{Boer:1997mf}
  D.~Boer, R.~Jakob and P.~J.~Mulders,
  Nucl.\ Phys.\  B {\bf 504}, 345 (1997).

\bibitem{Jaffe:1991kp}
  R.~L.~Jaffe and X.~D.~Ji,
  Phys.\ Rev.\ Lett.\  {\bf 67}, 552 (1991);
  Nucl.\ Phys.\  B {\bf 375}, 527 (1992).


\bibitem{Efremov:2006qm}
  A.~V.~Efremov, K.~Goeke and P.~Schweitzer,
  Phys.\ Rev.\  D {\bf 73}, 094025 (2006).

\bibitem{Anselmino:2007fs}
  M.~Anselmino, {\it et al.},
  Phys.\ Rev.\  D {\bf 75}, 054032 (2007).



\bibitem{hermes}
A.~Airapetian {\it et al.},
Phys.\ Rev.\ Lett.\  {\bf 84}, 4047 (2000);
  Phys.\ Rev.\ Lett.\  {\bf 94}, 012002 (2005);
  V.~Y.~Alexakhin {\it et al.},
  Phys.\ Rev.\ Lett.\  {\bf 94}, 202002 (2005).



\bibitem{belle}
  K.~Abe {\it et al.},
  Phys.\ Rev.\ Lett.\  {\bf 96}, 232002 (2006);
  Phys.\ Rev.\  D {\bf 78}, 032011 (2008)

\bibitem{Metz:2002iz}
  A.~Metz,
  Phys.\ Lett.\  B {\bf 549}, 139 (2002).

\bibitem{Gamberg:2008yt}
  L.~P.~Gamberg, A.~Mukherjee and P.~J.~Mulders,
  Phys.\ Rev.\  D {\bf 77}, 114026 (2008).

\bibitem{Meissner:2008yf}
  S.~Meissner and A.~Metz,
  arXiv:0812.3783 [hep-ph].


\bibitem{Koike:2002ti}
  Y.~Koike,
  AIP Conf.\ Proc.\  {\bf 675}, 449 (2003).


\bibitem{Ji:1993vw}
  X.~D.~Ji,
  Phys.\ Rev.\  D {\bf 49}, 114 (1994).


\bibitem{Ellis:1982wd}
  R.~K.~Ellis, W.~Furmanski and R.~Petronzio,
  Nucl.\ Phys.\  B {\bf 207}, 1 (1982).



\bibitem{meng}
  R.~Meng, F.~I.~Olness and D.~E.~Soper,
  Phys.\ Rev.\ D {\bf 54}, 1919 (1996).






\end{thebibliography}
\end{document}